# Certified Exact Transcendental Real Number Computation in Coq[*]

by Russell O'Connor

Institute for Computing and Information Science
Faculty of Science
Radboud University Nijmegen

*Email:* r.oconnor@cs.ru.nl

**Abstract**

Reasoning about real number expressions in a proof assistant is challenging. Several problems in theorem proving can be solved by using *exact* real number computation. I have implemented a library for reasoning and computing with complete metric spaces in the Coq proof assistant and used this library to build a constructive real number implementation including elementary real number functions and proofs of correctness. Using this library, I have created a tactic that automatically proves strict inequalities over closed elementary real number expressions by computation.

## 0 Licence

## 1 Introduction

Mathematics increasingly relies on computation for proofs. Because software is often error prone, proofs depending on computation are sometimes considered suspect. Recently, people have used proof assistants to verify these kinds of mathematical theorems [7]. Real number computation plays an essential role in some of these problems. These proofs typically require finding a rational approximation of some real number expression to within a specified error or proving a (strict) inequality between two real number expressions. Two examples of such proofs are the disproof of Merten's conjecture [15] and the proof of Kepler's conjecture [8]. Certified real number computation also has other applications including verifying properties of hybrid automata.

Proof assistants based on dependent type theory, such as Coq [17], allow one to develop a constructive theory of real numbers in which approximations of real numbers can be evaluated by the system. Functions on real numbers compute what accuracy is needed from their input to satisfy the requested accuracy for their output. Rather than accumulating rounding errors, the resulting approximations are guaranteed to be within the accuracy requested. One can develop a constructive theory of real numbers that yields efficient functions by taking care to ensure the computational aspects of the proofs are efficient. This paper illustrates how to develop such an efficient constructive theory. We begin reviewing some results that are detailed in a previous publication [14]:

- A theory of metric spaces is developed (Section 3) that is independent of the real numbers. An operation for completing metric spaces is defined (Section 3.2), and this operation is seen to be a monad.
- This theory of complete metric spaces is used to define the real numbers (Section 4). A key idea is to first define elementary functions over the rational numbers, and then, once the functions are shown to be uniformly continuous, lift these functions to the real numbers by using the monad operations.

---

A large library of mathematical results called CoRN has previously been developed at Radboud University Nijmegen [3]. Its collection of proofs includes both the fundamental theorem of algebra and the fundamental theorem of calculus. I extended this library by formalizing this theory of complete metric spaces. The new results detailing how this theory was formalized in Coq are covered (Section 5):

- The formalization was designed with efficient execution in mind (Section 5.1).
- Care was needed to efficiently approximate infinite series (Section 5.2).
- The technique of proof by reflection is used to verify a definition of $\pi$ (Section 5.3).
- Elementary functions are proved correct by showing that they are equivalent to their corresponding functions defined in the CoRN library (Section 5.4).
- This theory is put to use by developing a tactic that uses computation to automatically verify strict inequalities over closed real number expressions (Section 5.5).

This formalization will be part of the next version of the CoRN library, which will be released at the same time Coq 8.2 is released.

## 1.1 Notation

The propositions true and false are denoted by $\top$ and $\bot$ respectively. The type of propositions is written as $\star$. In Coq this type is `Prop`.

The type $\mathbb{Q}^+$ denotes the strictly positive rational numbers, and I will use similar notation for other number types. The type $\mathbb{Q}^+_\infty$ denotes $\mathbb{Q}^+ + \{\infty\}$.

Functions taking multiple arguments will be curried as in $f\colon A \Rightarrow B \Rightarrow C$; however, for readability, I will often use mathematical notation when applying parameters, $f(x, y)$, even though it should technically be written as $f(x)(y)$.

I denote the function $f$ iterated $n$ times as $f^{(n)}$.

Because constructive mathematics has a classical interpretation, all the theorems in this paper can also be understood as theorems of classical analysis. Although some of the definitions I use are somewhat different from the usual classical definitions, they are still equivalent (under classical logic) to their classical counterparts.

## 2 Background

The real numbers are typically defined as a Cauchy sequence of rational numbers. A sequence $x\colon \mathbb{N} \Rightarrow \mathbb{Q}$ is Cauchy when

$$\forall \varepsilon\colon \mathbb{Q}.\, 0 < \varepsilon \Rightarrow \exists N\colon \mathbb{N}.\, \forall m\colon \mathbb{N}.\, N \leq m \Rightarrow |x_m - x_N| \leq \varepsilon.$$

The function mapping $\varepsilon$ to $N$ is the modulus of convergence. It tells you how far into the sequence you must reach in order to get good rational approximations to the real number that $x$ represents.

By using the constructive existential, one ensures that the value of $N$ is computable from $\varepsilon$. This results in the constructive real numbers. One can compute approximations of constructive real numbers to within any given precision.

Real numbers are usually created from Cauchy sequences (which often arise from Taylor series). Perhaps this is why the Cauchy sequence definition is common. On the other hand, approximation is the fundamental operation for *consuming* real numbers. This suggests an alternative definition of real numbers based on how they are consumed. One can define a real number as a regular function of rational numbers. A regular function of rational numbers is a function $x\colon \mathbb{Q}^+ \Rightarrow \mathbb{Q}$ such that

$$\forall \varepsilon_1\, \varepsilon_2.\, |x(\varepsilon_1) - x(\varepsilon_2)| \leq \varepsilon_1 + \varepsilon_2.$$

Regular functions are a generalization of regular sequences, which Bishop and Bridges use to define the real numbers [1]. With regular functions, $x$ directly represents the function that approximates a real number to within $\varepsilon$. The regularity condition ensures that the approximations are coherent.

Regular functions and Cauchy sequences can be used to construct more than just the real numbers. They can be used to construct the completion of any metric space.

## 3 Metric Spaces

Usually a metric space $X$ is defined by a metric function $d\colon X \times X \Rightarrow \mathbb{R}$; however, this assumes that the real numbers have already been defined. Instead, one can define a metric space based on a ball relation $\beta_\varepsilon(a,b)$, that characterizes when $d(a,b) \leq \varepsilon$. Partial application, $\beta_\varepsilon(a)$, yields a predicate that represents the set of points inside the closed ball of radius $\varepsilon$ around $a$. The following axioms characterize a ball relationship $\beta\colon \mathbb{Q}^+ \Rightarrow X \Rightarrow X \Rightarrow \star$.

1. $\beta_\varepsilon(a,a)$
2. $\beta_\varepsilon(a,b) \Rightarrow \beta_\varepsilon(b,a)$
3. $\beta_{\varepsilon_1}(a,b) \Rightarrow \beta_{\varepsilon_2}(b,c) \Rightarrow \beta_{\varepsilon_1+\varepsilon_2}(a,c)$
4. $(\forall \delta\colon \mathbb{Q}^+.\, \beta_{\varepsilon+\delta}(a,b)) \Rightarrow \beta_\varepsilon(a,b)$

Axioms 1 and 2 state that the ball relationship is reflexive and symmetric. Axiom 3 is a form of the triangle inequality. Axiom 4 states that the balls are closed. Closed balls are used because their proof objects usually have no computational content and can be ignored during evaluation. For some metric spaces, such as the real numbers, open balls are defined with existential quantifiers and their use would lead to unnecessary computation [4].

Two points are considered identical if they are arbitrarily close to each other.

$$(\forall \varepsilon.\, \beta_\varepsilon(a,b)) \Leftrightarrow a \asymp b$$

This can be considered either the definition of equivalence in $X$, or if $X$ comes with an equivalence relationship, then it can be considered a fifth axiom.

In Coq, a metric space $X$ is a dependent record containing

1. a type (called the carrier)
2. a ball relation on that type
3. a proof that this ball relation satisfies the the above axioms.

The second projection function $B$ returns the ball relation component of the metric space. I will write the metric space parameter in a superscript, as in $B^X$. I will not distinguish between a metric space and its carrier, so $X$ will denote either a metric space or its carrier depending on the context.

Sometimes an extended ball relation $\check{B}^X \colon \mathbb{Q}^+_\infty \Rightarrow X \Rightarrow X \Rightarrow \star$ will be used where $\check{B}^X_\infty(a,b)$ always holds and reduces to $B^X_\varepsilon(a,b)$ when $\varepsilon < \infty$.

### 3.1 Uniformly Continuous Functions

A uniformly continuous function allows one to approximate the output from an approximation of the input. The usual definition for a function $f\colon X \Rightarrow Y$ to be uniformly continuous is

$$\forall \varepsilon.\, \exists \delta.\, \forall a\, b.\, B^X_\delta(a,b) \Rightarrow B^Y_\varepsilon(f(a), f(b)).$$

The function mapping $\varepsilon$ to $\delta$ is what Bishop and Bridges [1] call the *modulus of continuity* and is denoted by $\mu_f$. (This is the inverse of what mathematicians usually call the modulus of continuity.)

It is advantageous to use a more general notion of modulus of continuity that can return $\infty$. This is used for bounded functions when the requested accuracy is wider than the bound on the function. For example, $\mu_{\sin}(\varepsilon) = \infty$ for $1 \leq \varepsilon$ because $\sin(x)(\varepsilon) = 0$ for all $x$. We also pull out the modulus of continuity in order to reason about it directly. Thus, we define a function $f\colon X \Rightarrow Y$ to be *uniformly continuous with modulus* $\mu_f \colon \mathbb{Q}^+ \Rightarrow \mathbb{Q}^+_\infty$ when

$$\forall a\, b\, \varepsilon.\, \check{B}^X_{\mu_f(\varepsilon)}(a,b) \Rightarrow B^Y_\varepsilon(f(a), f(b)).$$

In Coq, a uniformly continuous functions is a dependent record containing

1. a function $f$ between two metric spaces
2. a modulus of continuity for $f$
3. a proof that $f$ is uniformly continuous with the given modulus.

This means that $\mu$ is really the second projection function. Again, I will not distinguish between the uniformly continuous function $f$ and its actual function.

I will denote the type of uniformly continuous functions with the single bar arrow, as in $X \to Y$.

## 3.2 Complete Metric Spaces

We are now in a position to define regular functions over an arbitrary metric space $X$. A function $x : \mathbb{Q}_\infty^+ \Rightarrow X$ is a *regular function* when

$$\forall \varepsilon_1\, \varepsilon_2 : \mathbb{Q}^+. B^X_{\varepsilon_1+\varepsilon_2}(x(\varepsilon_1), x(\varepsilon_2)).$$

The function $x$ is allowed to return anything when given $\infty$.

Two regular functions are equivalent ($x \asymp y$) when their approximations are arbitrarily close to each other.

$$\forall \varepsilon_1\, \varepsilon_2 : \mathbb{Q}^+. B^X_{\varepsilon_1+\varepsilon_2}(x(\varepsilon_1), y(\varepsilon_2))$$

Thus, a regular function is a function that is equivalent to itself under this relation.

Regular functions form a metric space [14], $\mathfrak{C}(X)$, where the ball relation $B^{\mathfrak{C}(X)}_\varepsilon(x,y)$ is

$$\forall \delta_1\, \delta_2 : \mathbb{Q}^+. B^X_{\delta_1+\varepsilon+\delta_2}(x(\delta_1), y(\delta_2)).$$

This states that $x$ and $y$ are within $\varepsilon$ of each other when their approximations are almost within $\varepsilon$ of each other.

### 3.2.1 Completion is a Monad

The completion operator $\mathfrak{C}$ forms a monad in the category of metric spaces and uniformly continuous functions between them [14]. The injection of $X$ into $\mathfrak{C}(X)$ is unit : $X \to \mathfrak{C}(X)$. The proof that a complete metric space is complete yields join : $\mathfrak{C}(\mathfrak{C}(X)) \to \mathfrak{C}(X)$. The function map : $(X \to Y) \Rightarrow \mathfrak{C}(X) \to \mathfrak{C}(Y)$ lifts uniformly continuous functions to the complete space. Finally, bind : $(X \to \mathfrak{C}(Y)) \Rightarrow \mathfrak{C}(X) \to \mathfrak{C}(Y)$ is defined in terms of map and join in the usual way.

$$\begin{aligned}
\mathrm{unit}(a)(\varepsilon) &:= a \\
\mathrm{join}(x)(\varepsilon) &:= x\left(\frac{\varepsilon}{2}\right)\left(\frac{\varepsilon}{2}\right) \\
\mathrm{map}(f)(x)(\varepsilon) &:= f\left(x\left(\frac{\widetilde{\mu_f}(\varepsilon)}{2}\right)\right) \\
\mathrm{bind}(f) &:= \mathrm{join} \circ \mathrm{map}(f)
\end{aligned} \qquad (1)$$

Here the function $\widetilde{\mu_f} : \mathbb{Q}_\infty^+ \Rightarrow \mathbb{Q}_\infty^+$ maps $\infty$ to $\infty$, and applies $\mu_f$ otherwise.

In my previous work, I used a simpler definition of map

$$\mathrm{map}'(f)(x)(\varepsilon) := f(x(\widetilde{\mu_f}(\varepsilon))). \qquad (2)$$

Unfortunately, this definition requires the additional assumption that $X$ be a prelength space [14]. Recently, I inferred from Richman's work [16] that map can be defined using equation 3.2.1 and works for all metric spaces if the modulus of continuity of map($f$) is smaller than $\mu_f$.

Despite the above, in the common case that $X$ is a prelength space, the definition of map' in equation 2 is more efficient, and map'($f$) has the same modulus of continuity as $f$. Because of this, I use map' (and similarly bind') throughout my work. I use map mostly for theoretical results.

### 3.2.2 Completion is a Strong Monad

Functions between two metric spaces form a metric space under the sup-norm. The ball relation between two functions $B_\varepsilon^{X \to Y}(f, g)$ is

$$\forall a. B_\varepsilon^Y(f(a), g(a))$$

Now the function $\mathrm{map} \colon (X \to Y) \to \mathfrak{C}(X) \to \mathfrak{C}(Y)$ can be shown to be uniformly continuous [14]. By defining $\mathrm{ap} \colon \mathfrak{C}(X \to Y) \to \mathfrak{C}(X) \to \mathfrak{C}(Y)$, higher arity maps such as $\mathrm{map2} \colon (X \to Y \to Z) \to \mathfrak{C}(X) \to \mathfrak{C}(Y) \to \mathfrak{C}(Z)$ can be constructed.

$$\begin{aligned} \mathrm{ap}(f)(x)(\varepsilon) &:= \mathrm{map}\Big(f\Big(\frac{\varepsilon}{2}\Big)\Big)(x)\Big(\frac{\varepsilon}{2}\Big) \\ \mathrm{map2}(f) &:= \mathrm{ap} \circ \mathrm{map}(f) \end{aligned}$$

## 4 Real Numbers

Because the rational numbers $\mathbb{Q}$ are a metric space, the real numbers can be simply defined as the completion of $\mathbb{Q}$.

$$\mathbb{R} := \mathfrak{C}(\mathbb{Q})$$

Uniformly continuous operations on the real numbers are defined by lifting their rational counterparts with map or map2. This is how $+_\mathbb{R}$ and $-_\mathbb{R}$ are defined [14].

I find using monadic operators to define functions on $\mathbb{R}$ is easier than trying to define functions directly. It splits the problem into two parts. The first part is to define the the function over $\mathbb{Q}$, which is easier to work with because equality is decidable for $\mathbb{Q}$. The second part is to prove that the function is uniformly continuous.

### 4.1 Order

A real number $x$ is non-negative when

$$\forall \varepsilon \colon \mathbb{Q}^+. -\varepsilon \leq_\mathbb{Q} x(\varepsilon).$$

The not-greater-than relation on real numbers, $x \leq_\mathbb{R} y$, means that $y - x$ is non-negative.

A real number $x$ is positive when

$$\exists \varepsilon \colon \mathbb{Q}^+. \mathrm{unit}(\varepsilon) \leq_\mathbb{R} x$$

(recall that $\mathrm{unit} \colon \mathbb{Q} \to \mathbb{R}$). One real number is less than another, $x <_\mathbb{R} y$, when $y - x$ is positive. Two real numbers are apart, $x \lessgtr y$, when $x < y \lor y < x$.

This definition of positivity differs from what would be analogous to Bishop and Bridges's definition, $\exists \varepsilon \colon \mathbb{Q}^+. \varepsilon <_\mathbb{Q} x(\varepsilon)$. Although the two definitions are equivalent, my definition above contains a rational number in $]0, x]$. This is exactly the information that will be needed to compute $x^{-1}$ or $\ln(x)$ (Section 4.2). With Bishop and Bridges's definition, one must compute $x(\varepsilon) - \varepsilon$, which is a potentially expensive calculation.

### 4.2 Non-uniformly Continuous and Partial Functions

Unfortunately not all functions that we want to consider are uniformly continuous. One can deal with continuous functions by noting that they are uniformly continuous on some collection of closed sub-domains that cover the whole space. For example, $\lambda a \colon \mathbb{Q}. a^2$ is uniformly continuous on $[-c, c]$. Thus, a real number $x$ can be squared by finding some domain $[-c, c]$ containing it and lifting $(\lambda a. (\max\ (\min\ (a,\ c),\ -c))^2$, which *is* uniformly continuous. In this case $c$ can be chosen to be $|x(1)| + 1$. One can prove that the result is independent of the choice of $c$, so long as $x \in [-c, c]$.

Evaluating a non-uniformly continuous function is potentially a costly operation. The input $x$ must be approximated twice. The first approximation finds a domain to operate in, and the second approximation is used to evaluate the function. In practice, I have found that one often has a suitable domain lying around for the particular problem at hand. If that is the case, then $x$ only needs to be approximated once.

Partial functions with open domains are handled in the same way as non-uniformly continuous functions. For example, $\lambda x.x^{-1}$ is uniformly continuous on the domains $[c, \infty[$ and $]-\infty, -c]$ (where $0 < c$). One difference is that one cannot automatically find a domain containing $x$. One requires a proof that $x$ is apart from 0. From such a proof, one can find a suitable domain containing $x$.

Partial functions with closed domains, such as $\lambda x. \sqrt{x}$, can be extended to continuous total functions. I extend the square root function to return 0 for negative values. If one wishes, one can then restrict the lifted function to only accept non-negative inputs.

### 4.3 Transcendental Functions

Transcendental functions are first defined from $\mathbb{Q}$ to $\mathbb{R}$. Once these functions are shown to be uniformly continuous (or otherwise using the techniques from the previous section), they are then lifted using bind to create functions from $\mathbb{R}$ to $\mathbb{R}$.

Most elementary functions can be defined on some sub-domain by an alternating decreasing series. Inputs outside this domain can often be dealt with by using range reduction. Range reduction uses elementary identities to reduce inputs from a wider to a narrower domain [14].

For example, the alternating series $\sum_{i=0}^{\infty} (-1)^i \frac{a^{2i+1}}{(2i+1)!}$ computes $\sin(a)$, and is decreasing when $a \in [-1, 1]$. For $a$ outside this interval, range reduction is preformed by repeated application of the identity

$$\sin(a) \asymp 3\sin\left(\frac{a}{3}\right) - 4\sin^3\left(\frac{a}{3}\right).$$

The value of an infinite alternating series, is represented by a regular function that finds a partial sum having an error no more than $\varepsilon$. When an alternating series is decreasing, finding such a partial sum is easy because the last term also represents the error. One only needs to accumulate terms until a term becomes less than $\varepsilon$.

Coq will not accept a general recursive function that computes the above partial sum. It requires a proof of termination. This is done by computing an upper bound on the number of terms that will be needed. Strategies for doing this efficiently in Coq are discussed Section 5.2.

The elementary functions, sin, cos, and $\tan^{-1}$ are defined as described in my previous publication [14]. The implementation of ln has been improved by defining it in terms of $\tanh^{-1}$,

$$\ln\left(\frac{n}{d}\right) := 2\tanh^{-1}\left(\frac{n-d}{n+d}\right).$$

However, the input is still range reduced into $[\frac{1}{2}, 2]$ before using the above formula.

I have also implemented a function to sum sub-geometric series (a series where $|a_{n+1}| \leq r|a_n|$). The error of the partial sums of these series is easy to compute from the last term and $r$. I now use this function to compute the $\exp(a)$ function for $a \in ]0, 1[$.

### 4.4 Compression

Without intervention, the numerators and denominators of rational numbers occurring in real number computations become too large for practical computation. To help prevent this, I defined a compression operation for real numbers.

$$\mathrm{compress}(x)(\varepsilon) := \mathrm{approx}_{\mathbb{Q}}\left(x\left(\frac{\varepsilon}{2}\right), \frac{\varepsilon}{2}\right)$$

where $\mathrm{approx}_{\mathbb{Q}}(a, \delta)$ returns some rational number within $\delta$ of $a$. The idea is that $\mathrm{approx}_{\mathbb{Q}}(a, \delta)$ quickly computes a rational number close to $a$ but having a smaller numerator and denominator. In my implementation, I return $\frac{b}{2^n}$, where $2^n$ is the smallest power of 2 greater than the denominator of $\delta$, and $b$ is chosen appropriately so that the result is within $\delta$ of $a$.

The compress function is equivalent to the identity function on $\mathbb{R}$.

$$\mathrm{compress}(x) \asymp x$$

By liberally inserting compress into one's expressions, one can often dramatically improve the efficiency of real number calculations. I am considering adding a call to compress with every use of map or bind so that the user does not need to add these calls themselves. Too many calls to compress can harm performance but perhaps not enough to cause worry.

## 5 Formalization in Coq

The theory of metric spaces and real numbers described in Sections 3 and 4 has been formalized in the Coq proof assistant. I developed functions and proofs simultaneously. I did not extract functions from constructive proofs, nor did I write functions entirely separately from their proofs of correctness. Proofs and functions are often mixed together, such as in the dependent records of metric spaces, uniformly continuous functions, and regular functions.

### 5.1 Efficient Proofs

A mixture of proofs and functions can still be efficient to evaluate by taking care to write the functional aspects efficiently and ensuring that the non-functional aspects are declared opaque. Declaring lemmas as opaque prevents call-by-value evaluation from normalizing irrelevant proofs.

I used Coq's `Prop`/`Set` distinction (two different universes of types) to assist in the separation of these concerns [4]. Types that have at most one member (extensionally) are proof-irrelevant and go into `Prop`. Lemmas having these types are declared opaque. Types that may have more than one member go into `Set`, and objects of such types are kept transparent. This criterion means that I use the `Set` based sum and dependent pair types for the constructive disjunction and constructive existential quantifier.

When proving a constructive existential goal, one has to deal with both `Prop` and `Set` during a proof. The existential lives in `Set`, but after supplying the witness, a `Prop` based proof obligation remains. The witness needs to be transparent, but the proof obligation should be opaque. It is best to try and separate these two parts into two different definitions, one transparent and one opaque. However, in some instances I make the entire development transparent, but I mark the proof obligation part with Coq's `abstract` tactic. The `abstract` tactic automatically defines an opaque lemma containing marked part of the proof and places this lemma into the proof object. Thus, the marked part is never evaluated.

### 5.2 Summing Series

One of the more challenging aspects of the formalization was computing the infinite series defined in Section 4.3 in an efficient manner. In order to convince Coq that the procedure of accumulating terms until the error becomes sufficiently small terminates, I provided Coq with an upper bound on the number of terms that would be required. I tried two different methods to accomplish this.

The first method computes an upper bound on the number of terms needed as a Peano natural number. The problem is that the call-by-value evaluation scheme used by Coq's virtual machine would first compute this value before computing the series. This upper bound is potentially extremely large, it is encoded in unary, and only a few terms may actually be needed in the computation. The solution to this problem was to create a lazy natural number using the standard trick of placing a function from the unit type inside the constructor.

```
Inductive LazyNat : Set :=
| LazyO : LazyNat
| LazyS : (unit -> LazyNat) -> LazyNat.
```

**Figure 1.** Inductive definition of lazy natural numbers

The lambda expressions inside the lazy natural numbers delay the evaluation of the call-by-value scheme. With some care, only the number of constructors needed for the recursion are evaluated.

A second method, suggested by Benjamin Grégoire, is to compute the number of terms needed as a binary number. This prevents the term from becoming too big. It is possible to do recursion over the binary natural numbers such that two recursive calls are made with the output of one recursive call being threaded through the other. In this way, up to $n$ recursive calls can be made even though only $\lg n$ constructors are provided by the witness of termination.

In the simplified example below, the function `F` is iterated up to `n` times. Continuation passing style is used to thread the recursive calls.

```
Variable A R : Type
Variable F : (A -> R) -> A -> R

Fixpoint iterate_pos (n:positive) (cont: A -> R) : A -> R :=
match n with
| xH => F cont
| xO n' => iterate_pos n' (fun a => iterate_pos n' cont a)
| xI n' => F (fun a => (iterate_pos n'
                         (fun a => iterate_pos n' cont a)) a)
end.
```

**Figure 2.** The Coq function `iterate_pos` recurses `F` at up to `n` times, using continuation passing style.

The $\eta$-expansion of the continuations in the above definition are important, otherwise the virtual machine would compute the value of the `iterate_pos n' cont` calls before reducing `F`. This is important because `F` may not utilize its recursive call depending on the value of `a`. In such a case, we do not want the recursive call to be evaluated.

### 5.3 $\pi$

A common definition of $\pi$ is $4\tan^{-1}(1)$. This is an inefficient way of computing $\pi$ because the series for $\tan^{-1}(1)$ converges slowly. One can more efficiently compute $\pi$ by calling $\tan^{-1}$ with smaller values [18]. I chose an optimized formula for $\pi$ from a list [19]:

$$\pi := 176\tan^{-1}\left(\frac{1}{57}\right) + 28\tan^{-1}\left(\frac{1}{239}\right) - 48\tan^{-1}\left(\frac{1}{682}\right) + 96\tan^{-1}\left(\frac{1}{12943}\right)$$

This formula can easily be shown to be equivalent to $4\tan^{-1}(1)$ by repeated application of the *arctangent sum law*:

$$\text{if } a,b \in ]-1,1[ \text{ then } \tan^{-1}(a) + \tan^{-1}(b) \asymp \tan^{-1}\left(\frac{a+b}{1-ab}\right)$$

To apply the arctangent sum law, one needs to verify that $a$ and $b$ lie in $]-1,1[$. To solve this, I wrote a Coq function to iterate the function $f(b) := \frac{a+b}{1-ab}$, and at each step verify that the result is in the interval $]-1,1[$. This function, called `ArcTan_multiple`, has type

$$\forall a: \mathbb{Q}. -1 < a < 1 \Rightarrow \forall n. \top \vee \left(n\tan^{-1}(x) \asymp \tan^{-1}(f^{(n)}(0))\right)$$

It is easy to build a function of the above type that just proves $\top$ in all cases, but `ArcTan_multiple` tries to prove the non-trivial result if it can.

To apply this lemma I use a technique called reflection. The idea is to evaluate the `ArcTan_multiple`$(a,r,n)$ into head normal form. This will yield either `left`$(q)$ or `right`$(p)$. If `right`$(p)$ is returned then $p$ is the proof we want.

My first attempt at building a tactic to implement this did not work well. I used Coq's `eval hnf` command to reduce my expression to head normal form. However, this command repeatedly calls `simpl` to expose a constructor instead of using the evaluation mechanism directly. The problem was that `simpl` does extra reductions that are not necessary to get head normal form, so using `eval hnf` was too time consuming.

Instead, I built a reflection lemma, called `reflect_right`, to assist in applying the `ArcTan_multiple` function:

$$\forall z: A \vee B.(\text{if } z \text{ then } \bot \text{ else } \top) \Rightarrow B$$

This simple lemma does case analysis on $z$. If $z$ contains a proof of $A$, it returns a proof of $\bot \Rightarrow B$. If $z$ contains a proof of $B$, it returns a proof of $\top \Rightarrow B$. To prove $n \tan^{-1}(a) \asymp \tan^{-1}(f^{(n)}(0))$, for the example $a := \frac{1}{57}$ and $n := 176$, one applies `reflect_right` composed with `ArcTan_multiple` to reduce the goal to

$$\texttt{if (ArcTan\_multiple } \tfrac{1}{57} * \texttt{ 176) then } \bot \texttt{ else } \top,$$

where $*$ is the trivial proof of $-1 < \frac{1}{57} < 1$. Then one normalizes this expression using *lazy evaluation* to either $\top$, if `ArcTan_multiple` succeeds, or $\bot$, if it fails.

## 5.4 Correctness

There are two ways to prove that functions are correct. One way is to prove that they satisfy some uniquely defining properties. The other way is to prove that the functions are equivalent to a given reference implementation. I have verified that my elementary functions are equivalent to the corresponding functions defined in the CoRN library [3]. The functions in the CoRN library can be seen to be correct from the large library of theorems available about them. The CoRN library contains many different characterizations of these functions and new characterizations can easily be developed.

The CoRN library defines a *real number structure* as a complete, ordered, Archimedean field. My first step was to prove that my operations form a real number structure. I first attempted to directly show that my real numbers satisfy all the axioms of a real number structure, but this approach was difficult. Instead, I created an isomorphism between my real numbers and the existing model of the real numbers developed by Niqui [6]. This was a much easier approach because Niqui's Cauchy sequence definition and my regular function definition are closely related. With this isomorphism in place, I proved my operations satisfied the axioms of a real number structure by passing through the isomorphism and using Niqui's existing lemmas. Niqui has also proved that all real number structures are isomorphic, so I can create an isomorphism between my real numbers and any other real number structure.

The next step was to define my elementary functions and prove that they are equivalent to the corresponding CoRN functions. These theorems are of the form $\Phi(f_{\text{CoRN}}(x)) \asymp f(\Phi(x))$ where $\Phi$ is the isomorphism from CoRN's real numbers to my real numbers.

To aid in converting statements between different representations of real numbers, I have created a rewrite database that contains the correctness lemmas. By rewriting with this database, expressions can be automatically converted from CoRN's real numbers into my real numbers. This database can easily be extended with more functions in the future.

The CoRN library was more than just a specification; this library was useful throughout my development. For example, I was often able to prove that a differentiable function $f$ is uniformly continuous with modulus $\lambda \varepsilon. \frac{\varepsilon}{M}$ when $M$ is a bound on the derivative of $f$. I could prove this because the theory of derivatives had already been developed in CoRN. The CoRN library also helped me reduce the problem of proving the correctness of continuous functions on $\mathbb{R}$ to proving correctness only on $\mathbb{Q}$.

## 5.5 Solving Strict Inequalities Automatically

Whether a strict inequality holds between real numbers is semi-decidable. This question can be reduced to proving that some expression $e_0 : \mathbb{R}$ is positive. To prove $e_0$ is positive one must find an $\varepsilon : \mathbb{Q}^+$, such that $\text{unit}(\varepsilon) \leq e_0$. I wrote a tactic to automate the search for such a witness. It starts with an initial $\delta : \mathbb{Q}^+$, and computes to see if $e_0(\delta) - \delta$ is positive. If it is positive, then $e_0(\delta) - \delta$ is such a witness; otherwise $\delta$ is halved and the process is repeated. If $e_0 \asymp 0$, then this process will never terminate. If $e_0 < 0$, then the tactic will notice that $e_0(\delta) + \delta$ is negative and terminate with an error indicating that $e_0$ is negative.

This tactic has been combined with the rewrite database of correctness lemmas to produce a tactic that solves strict inequalities of closed expressions over CoRN's real numbers. This allows users to work entirely with CoRN's real numbers. They need never be aware that my effective real numbers are running behind the scenes.

Recently Cezary Kaliszyk has proved that Coq's classical real numbers (from the standard library) form a CoRN real number structure, and he has shown that Coq's elementary functions are equivalent to CoRN's. Now strict inequalities composed from elementary functions over Coq's classical real numbers can automatically be solved.

The tactic currently only works for expressions composed from total functions. Partial functions with open domains pose a problem because proof objects witnessing, for example, that $x$ is positive for $\ln(x)$ must be transparent for computation. However, proof objects for CoRN functions are opaque, and Coq's classical functions have no proof objects. The required proof objects are proofs of strict inequalities, so I am developing a tactic that recursively solves these strict inequalities and creates transparent proof objects. This will allow one prove strict inequalities over expressions that include partial functions such as ln and $\lambda x.x^{-1}$.

## 5.6 Setoids

Coq does not have quotient types. Setoids are used in place of quotient types. A *setoid* is a type associated with an equivalence relation on that type. A framework for working with setoids is built into Coq. Coq allows one to associate an equivalence relation with a type and register functions as morphisms by proving they are well-defined with respect to the given equivalence relations. Coq allows you substitute terms with other equivalent terms in expressions composed from morphisms. Coq automatically creates proof objects validating these substitutions.

Setoids have some advantages over quotient types. Some functions, most notably the function that approximates real numbers, are not well-defined with respect to the equivalence relation—two equivalent real numbers may compute different approximations. It is unclear how one would support these functions if a system with quotient types was used.

Support for setoids was invaluable for development; however, I encountered some difficulties when dealing with convertible types. The types `CR`, `Complete Q_as_MetricSpace`, and `cs_crr CRasCRing`, where `cs_crr` retrieves the carrier type, are all convertible. They are equivalent as far as the underlying type theory is concerned, but Coq's tactics work on the meta-level where these terms are distinguishable. The setoid system does not associate the equivalence relation on the real numbers with all of these various forms of the same type. Adding type annotations was not sufficient; they were simplified away by Coq. Instead, I used an identity function to force the types into a suitable form:

```
Definition ms_id (m:MetricSpace) (x:m) : m := x.
```

The setoid system is being reimplemented in the upcoming Coq 8.2 release. Therefore, some of these issues may no longer apply.

## 5.7 Timings

Table 1 shows examples of real number expressions that can be approximated. Approximations of these expressions were evaluated to within $10^{-20}$ on a 1.4 GHz ThinkPad X40 laptop using Coq's `vm_compute` command for computing with its virtual machine. These examples are taking from the "Many Digits" friendly competition problem set [13].

| Coq Expression | | | |
|---|---|---|---|
| **Mathematical Expression** | **Time** | **Result** | **Error** |
| `(CRsqrt (compress (rational_exp (1))*compress (CRinv_pos (3#1) CRpi)))%CR` | | | |
| $\sqrt{\frac{e}{\pi}}$ | 1 sec | 0.93019136710263285866 | $10^{-20}$ |
| `(sin (compress (CRpower_positive 3` `(translate (1#1) (compress (rational_exp (1))))))))%CR` | | | |
| $\sin((e+1)^3)$ | 25 sec | 0.90949524105726624718 | $10^{-20}$ |
| `(exp (compress (exp (compress (rational_exp (1#2))))))%CR` | | | |
| $e^{e^{e^{\frac{1}{2}}}}$ | 146 sec | 181.33130360854569351505 | $10^{-20}$ |

**Table 1.** Timings of approximations of various real number expressions.

## 6 Related Work

Julien is developing an implementation of real numbers in Coq using co-inductive streams of digits [11]. This representation allows common subexpressions to be easily shared because streams naturally memoize. Sharing does not work as well with my representation because real numbers are represented by functions. One would require additional structure to reuse approximations between subexpressions. Julien also uses the new machine integers implementation in Coq's virtual machine to make his computations even faster. It remains to be seen if using machine integers would provide a similar boost in my implementation.

Cruz-Filipe implemented CoRN's library of theorems and functions over the real numbers in Coq [2]. His implementation forms the reference specification of my work. Although his implementation is constructive, it was never designed for evaluation [5]. Many important definitions are opaque and efficiency of computation was not a concern during development. Cruz-Filipe showed that it is practical to develop a constructive theory of real analysis inside Coq. My work extends this result to show that it is also possible to develop a theory of real analysis that is practical to evaluate.

Muñoz and Lester implemented a system for approximating real number expressions in PVS [12]. Their system uses rational interval analysis for doing computation on monotone segments of transcendental functions. Unfortunately, this leads to some difficulties when reasoning at a local minimum or maximum, so their system cannot automatically prove $0 < \sin\left(\frac{\pi}{2}\right)$, for instance.

Harrison implemented a system to approximate real number expressions in HOL Light [9]. His system runs a tactic that externally computes an approximation to an expression and generates a proof that the approximation is correct. If such a technique were implemented for Coq, it would generate large proof objects. This is not an issue in HOL Light where proof objects are not kept.

Jones created a preliminary implementation of real numbers and complete metric spaces in LEGO [10]. She represented real numbers as a collection containing arbitrarily small intervals of rational numbers that all intersect. Complete metric spaces were similarly represented by using balls in place of intervals. Because the only way of getting an interval from the collection is by using the arbitrarily small interval property, her representation could have been simplified by removing the collection and let it implicitly be the image of a function that produces arbitrarily small intervals. This is similar to my work because one can interpret a regular function $f$ as producing the interval $[f(\varepsilon) - \varepsilon, f(\varepsilon) + \varepsilon]$. Perhaps using functions that return intervals could improve computation by allowing one to see that an approximation maybe more accurate than requested.

My work is largely based on Bishop and Bridges's work [1]. Some definitions have been modified to make the resulting functions more efficient. My definition of a metric space is more general; it does not require that the distance function be computable. The original motivation for the ball relation was only to develop a theory of metric spaces that did not presuppose the existence of the real numbers; however, it allows me to form a metric space of functions. This metric space does not have a computable distance function in general and would not be a metric space according to Bishop and Bridge's definition.

## 7 Conclusion

We have seen a novel definition of a metric space using a ball relation. We have seen how to create an effective representation for complete metric spaces and seen that the completion operation forms a monad. Using this monad, we defined the real numbers and used the monad operations to define effective functions on the real numbers. This theory has been formalized in Coq, and the elementary functions have been proved correct. Real number expressions can be approximated to any precision by evaluation inside Coq. Finally, a tactic was developed to automatically proof strict inequalities over closed real number expressions.

After completing the Haskell prototype and after writing up detailed paper proofs [14], it took about five months of work to complete the Coq formalization. This preparation allowed for a smooth formalization experience. Only a few minor errors were found in the paper proofs. These errors mostly consisted of failing to consider cases when $\varepsilon$ may be too large, and they were easy to resolve.

My results show that one can implement constructive mathematics such that the resulting functionally can be efficiently executed. This may be seen as the beginning of the realization of Bishop's program to see constructive mathematics as programming language.